\documentclass[twocolumn,showpacs,preprintnumbers,amsmath,amssymb,floatfix]{revtex4}

\usepackage{graphicx}
\usepackage{dcolumn}
\usepackage{bm}

\newcommand{\beq}{\begin{equation}}
\newcommand{\eeq}{\end{equation}}
\newcommand{\bey}{\begin{eqnarray}}
\newcommand{\eey}{\end{eqnarray}}

\begin{document}

\title{ Geodesic study of a charged black hole}

\author{ Mehedi Kalam}
\email{kalam@iucaa.ernet.in} \affiliation{Department of
Physics, Aliah University, Sector - V , Salt Lake,  Kolkata -
700091, India}
\author{Nur Farhad}
\email{farhad.nur@gmail.com} \affiliation{Department of Mathematics,
Aliah University, Sector - V , Salt Lake,  Kolkata, India}
\author{Sk. Monowar Hossein}
\email{sami_milu@yahoo.co.uk} \affiliation{Department of
Mathematics, Aliah University, Sector - V , Salt Lake,  Kolkata -
700091, India}

\date{\today}

\begin{abstract}
The behavior of the timelike and null geodesics of charged E.
Ay$\acute{o}$n-Beato and A. Garcia (ABG) black hole are
investigated. For circular and radial geodesics, we investigate
all the possible motions by plotting the effective potentials for
different parameters. In conclusion, we have shown that there is
no phenomenon of \textit{superradiance} in this case.
\end{abstract}


 \maketitle

\section{Introduction}

   A black hole can be characterized by its mass M, charge Q and
   angular momentum J. A rotating charged black hole is
   represented by Kerr-Newman metric while uncharged one can be
   represented as Kerr black hole. Again, if a non-rotating black
   hole has charged then it is called Reissner-Nordstr\"{o}m
   black hole. For uncharged case, the black hole which has only
   mass dependency is called Schwarzschild black hole. \\

   Recently, E.
Ay$\acute{o}$n-Beato and A. Garcia \cite{Beato1999} (ABG) gave  a
solution of  the Einstein field equations with nonlinear
electrodynamics, known as ABG black hole.

 The static spherical symmetric  space-time  is
described by the metric
\begin{equation}
ds^2 = -f(r)dt^2 +  \frac{1}{f(r)}dr^2 +r^2 (d\theta^2
+sin^2\theta d\phi^2), \label{eq1}
\end{equation}
where  $~~f(r)=1-\frac{2M}{r}+\frac{2M}{r}\tanh(\frac{Q^2}{2Mr})$\\
with M and Q corresponding to Mass,electro-magnetic charge of the
black hole respectively.\\

This solution corresponds to a regular black hole with mass M and
electro-magnetic charge Q and avoids thus the singularity problem.
Also, the metric  behaves asymptotically as the
Reissner-Nordstr\"{o}m (RN)black hole solution. It is clear that
the singularity of the RN solution, at r = 0, has been omitted
and now it simply becomes the origin of the spherical
coordinates.Several \cite{Kalam2009, Raychaudhuri2009} authors
studied different types of black holes (like charged brane-world
black holes, dyadosphere of a charged black hole etc.). I.
Radinschi \cite{Radinschi2000} calculate the energy distribution
of a charged ABG black hole by using the energy-momentum
complexes of
Einstein and M\o ller.\\

In this paper, we will analyze the behaviour of the timelike and
null geodesics of charged E.Ay$\acute{o}$n-Beato and A. Garcia
(ABG) black hole. For circular and radial geodesics, we will
discuss the possible motions by plotting the effective potentials
for various parameters. We also check the phenomenon of
superradiance for an incident massless scalar field for such a
black hole.

\section{The Geodesics}

 Now, the equation for the geodesics in the metric (1) is given
\begin{equation}
\frac{d^2x^\mu}{d\tau^{2}}+\Gamma^{\mu}_{\sigma\lambda}\frac{dx^\sigma}{d\tau}\frac{dx^\lambda}{d\tau}=0
\end{equation}
From eqn.(2) we have
\begin{eqnarray}
\frac{1}{f(r)}(\frac{dr}{d\tau})^2 &=& \frac{E^2}{f(r)}-\frac{J^2}{r^2}-L \nonumber \\
r^2(\frac{d\phi}{d\tau})&=& J \nonumber \\
\frac{dt}{d\tau}&=& \frac{E}{f(r)}
\end{eqnarray}
where we assume the$~\theta=\frac{\pi}{2}$ plane and constants E and J are the energy per unit mass and angular momentum,respectively about an axis perpendicular to the invariant plane $\theta=\frac{\pi}{2}$. Here, the affine parameter is $\tau$ and the Lagrangian,L has values 0 and 1,respectively for massless and massive particles. \\

 Now, for radial geodesic (J=0)
\begin{equation}
\dot{r}^2 \equiv \left(\frac{dr}{d\tau}\right)^2=E^2 - Lf(r)
\end{equation}
Using the above equations we get
\begin{eqnarray}
\left(\frac{dr}{dt}\right)^2&=&\left(1-\frac{2M}{r}+\frac{2M}{r}\tanh(\frac{Q^2}{2Mr})\right)^2\nonumber
\\&&\left[1-\frac{L}{E^2}\left(1-\frac{2M}{r}+\frac{2M}{r}\tanh(\frac{Q^2}{2Mr})\right)\right]
\end{eqnarray}
\subsection{Photonlike  particle motion(L=0)}
 For photonlike particle,we have
\begin{eqnarray}
\left(\frac{dr}{dt}\right)^2=\left[1-\frac{2M}{r}+\frac{2M}{r}\tanh(\frac{Q^2}{2Mr})\right]^2
\end{eqnarray}
i.e.
\begin{equation}
 \pm t=\int\frac{dr}{\left(1-\frac{2M}{r}+\frac{2M}{r}\tanh(\frac{Q^2}{2Mr})\right)}
\end{equation}

Considering, $1-\tanh(z)~\approx~z~$,~for a certain nbd. of z where $\frac{Q^2}{2Mr}=z$, we get
\begin{eqnarray}
\pm~t~=~\frac{Q^2}{2Mz}~-~\frac{Q}{2}\ln\left|\frac{Q+2Mz}{Q-2Mz}\right|\nonumber
\end{eqnarray}

i.e.\\
\begin{equation}
~\pm~t~=~r-\frac{Q}{2}\ln\left|\frac{Qr+Q^2}{Qr-Q^2}\right|
\end{equation}
The relation between time and distance for this particle is shown in the Fig. 1.

Now,we get from eqn.(4) as
\begin{equation}
\dot{r}^2 \equiv \left(\frac{dr}{d\tau}\right)^2=E^2 \nonumber
\end{equation}
that implies the $\tau~ - ~r$ relationship as
\begin{equation}
 \pm E \tau =r \nonumber
\end{equation}
The variation of proper time ($\tau$) with respect to the radial distance.(r).is shown in Fig.2.

\begin{figure}[htbp]
    \centering
        \includegraphics[scale=.3]{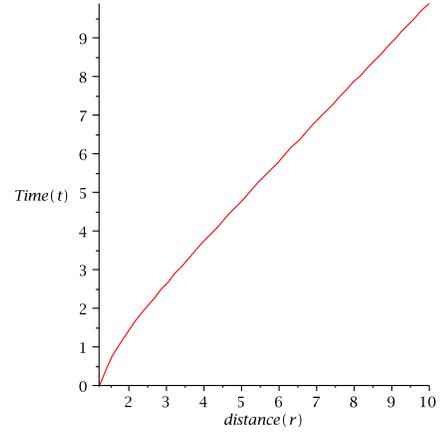}
        \caption{ Time-Distance relationship for photonlike particle(taking Q=1).}
    \label{fig:1}
\end{figure}

\subsection{ Massive particle motion (L=1)}

\begin{figure}[htbp]
    \centering
        \includegraphics[scale=.3]{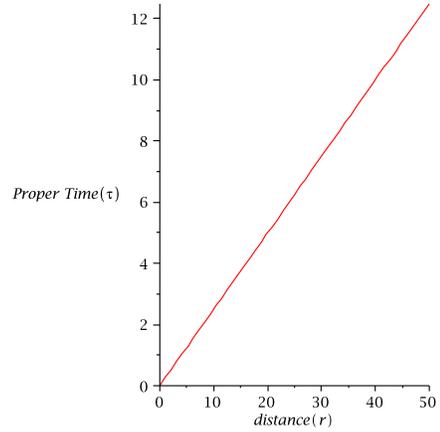}
       \caption{ Propertime-Distance relationship for photonlike particle(taking E=4).}
    \label{fig:2}
\end{figure}

For this case we can write,
\begin{eqnarray}
 \left(\frac{dr}{dt}\right)^2&=&\frac{1}{E^2}\left(1-\frac{2M}{r}+\frac{2M}{r}\tanh(\frac{Q^2}{2Mr})\right)^2\nonumber
\\&&\left[E^2-1+\frac{2M}{r}~-~\frac{2M}{r}\tanh(\frac{Q^2}{2Mr})\right] \nonumber
\end{eqnarray}
Assuming $\frac{Q^2}{2Mr}~=~z$ we get

\begin{eqnarray}
\pm~t= -\frac{Q^2E}{2M}\nonumber
\\\int\frac{dz}{z^2(1-\frac{4M^2}{Q^2}z(1-\tanh z))\sqrt{E^2-1+\frac{4M^2}{Q^2}z(1-\tanh z)}}
\end{eqnarray}
Again considering, $1-\tanh(z)~\approx~z~$,~for a certain nbd. of z where $\frac{Q^2}{2Mr}=z$, we get
\begin{equation}
\pm~t= -\frac{Q^2E}{2M}\nonumber
\\\int\frac{dz}{z^2(1-\frac{4M^2}{Q^2}z^2)\sqrt{E^2-1+\frac{4M^2}{Q^2}z^2}}
\end{equation}

Again asssuming,  $u~=~\frac{2M}{Q}z$ we get

\begin{equation}
\pm~t~=~-~\frac{QE}{\sqrt{E^2-1}}\int\frac{du}{u^2\left(1-u^2\right)\sqrt{1+\frac{u^2}{E^2-1}}}
\end{equation}

After simplification we get
\begin{equation}
\pm~t~=~-~QE\int\frac{dq}{q^2(1-E^2q^2)}
\end{equation}
where $ q = sin \left(tan^{-1}\left(\frac{u}{\sqrt{E^2-1}}\right)\right)$\\

After integrating,
\begin{equation}
\pm~t=Er\sqrt{E^2-1+\frac{Q^2}{2r^2}}-\frac{QE^3}{2}\ln\left|\frac{\sqrt{E^2-1+\frac{Q^2}{r^2}}+E\frac{Q}{r}}{\sqrt{E^2-1+\frac{Q^2}{r^2}}-E\frac{Q}{r}}\right|
\end{equation}
This will give the time(t)-distance(r) relationship(Fig.3).\\
\begin{figure}[htbp]
    \centering
        \includegraphics[scale=.3]{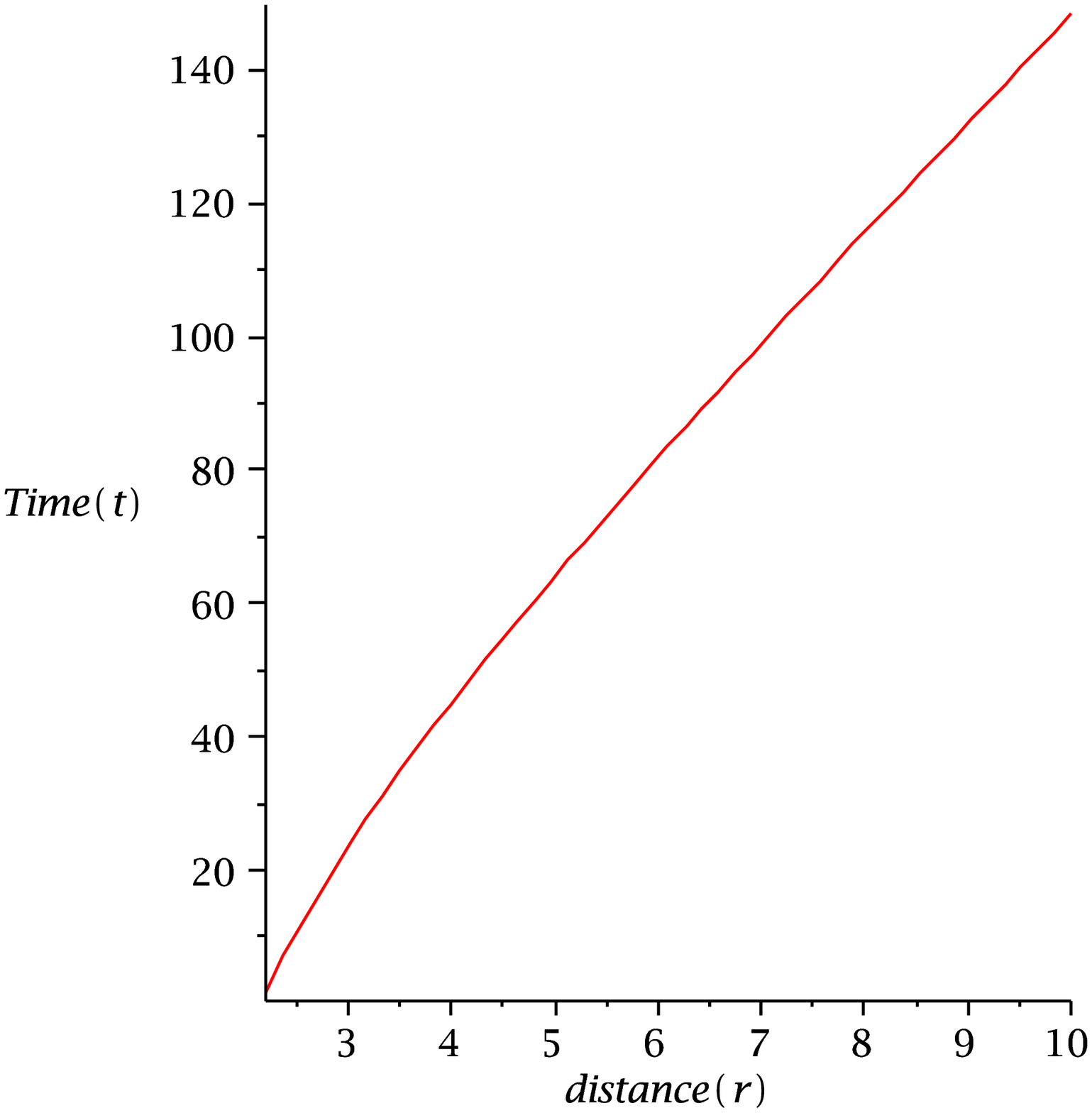}
        \caption{ Time-Distance relationship for massive particle(taking E=4,Q=1).}
    \label{fig:3}
\end{figure}
Again, from equation(4) we get
\begin{equation}
\dot{r}^2=\left(\frac{dr}{d\tau}\right)^2=E^2-\left(1-\frac{2M}{r}+\frac{2M}{r}\tanh(\frac{Q^2}{2Mr})\right)
\end{equation}
After simplification in a similar manner we get
\begin{equation}
 \pm\tau=~\frac{r}{(E^2-1)Q^2}\sqrt{E^2-1+\frac{Q^2}{r^2}}
\end{equation}
which gives the proper time($\tau$) -distance(r) relationship(Fig.4).

\begin{figure}[htbp]
    \centering
        \includegraphics[scale=.3]{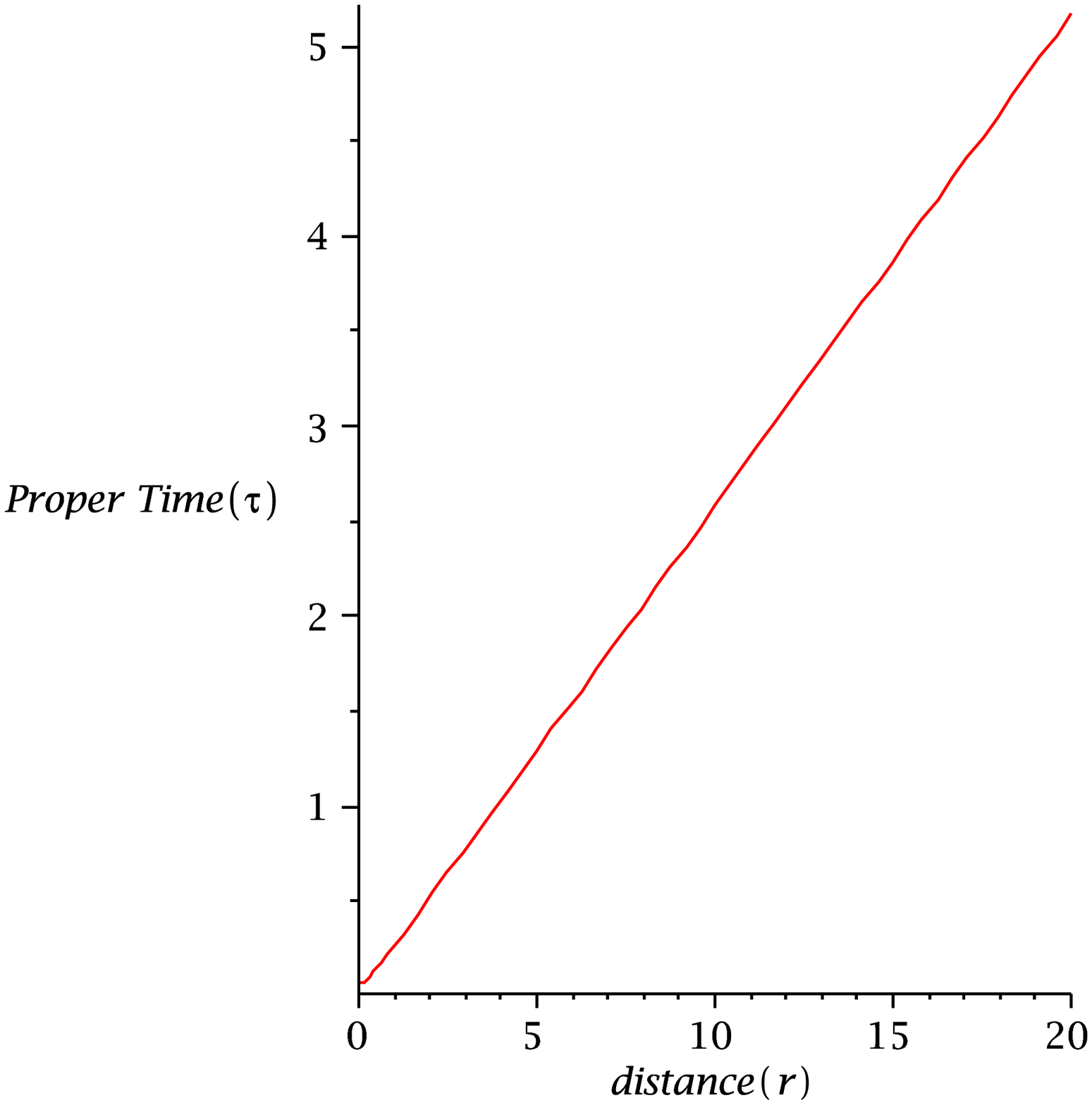}
        \caption{ Propertime-Distance relationship for massive particle(taking E=4,Q=1).}
    \label{fig:4}
\end{figure}

\section{Effective Potential}
From the geodesic eqn.(2) and (3) we get
\begin{equation}
\frac{1}{2}\left(\frac{dr}{d\tau}\right)^2=\frac{1}{2}\left[E^2~-~f(r)\left(\frac{J^2}{r^2}+L\right)\right]
\end{equation}

If we compare eqn.(15) with $\frac{\dot{r}^2}{2}+V_{eff}=0$, one can get the effective potential, which depends on E and L as follows :
\begin{equation}
V_{eff}=-\frac{1}{2}\left[E^2-f(r)\left(\frac{J^2}{r^2}+L\right)\right]
\end{equation}

\subsection{For Photonlike  particle(L=0)}
Consider, the radial geodesics where $J=0$. Then, the corresponding $V_{eff}$ is given by - \\
\begin{equation}
V_{eff}~=~-\frac{1}{2}E^2 \nonumber
\end{equation}

 The particle will behaves like a "free particle" i.e its $V_{eff}=0$ when energy,$E=0$ .\\

\begin{figure}[htbp]
    \centering
        \includegraphics[scale=.3]{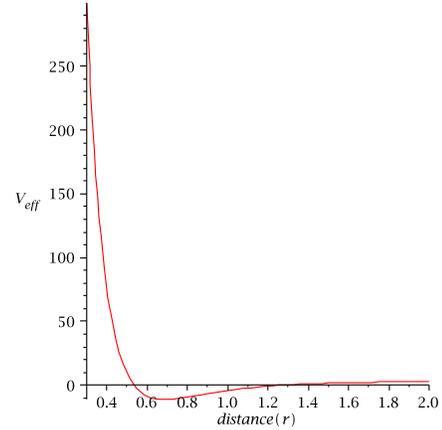}
        \caption{ Behaviour of the Effective Potential,$V_{eff}$ for ~~~~~~~~~~~~~~ $J\neq0$(taking E=1,J=10,Q=M=1)}
    \label{fig:5}
\end{figure}

We can easily say that the behavior of these geodesics is independent on the mass and charge of the black hole.
Now, for circular geodesics, $J~\neq~0$. The effective potential can be written as,
\begin{equation}
V_{eff}=-\frac{E^2}{2}+\frac{J^2}{2r^2}\left(1-\frac{2M}{r}+\frac{2M}{r}~\tanh \frac{Q^2}{2Mr}\right)
\end{equation}

If $r~\rightarrow0$, the effective potential , $V_{eff}(r)$ is large enough and approaches to  $-~\frac{E^2}{2}~$ when $r~\rightarrow~~\infty$ . At the horizon, the effective potential, $V_{eff}~=~-\frac{E^2}{2}~.$\\
We assume, the effective potential for $E~=~0$ [in eqn.(17), put $E~=~0$]. The roots of the potential is nothing but the horizons for this case. It is nice to see the effective potential,$V_{eff} $ is negative between the horizons. Hence, the particle would be bounded within the horizons. Again, a stable circular orbit will definitely exist between the horizons as $V_{eff}$ has a minima there(Fig.5).\\
\begin{figure}[htbp]
    \centering
        \includegraphics[scale=.3]{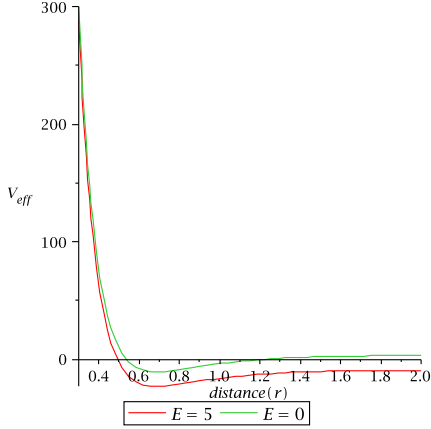}
        \caption{ Behaviour of the Effective Potential,$V_{eff}$ (for radial geodesic) for massive particle(taking  Q=M=1).}
    \label{fig:6}
\end{figure}
\subsection{For massive  particle(L=1)}
The effective potential is given by
\begin{equation}
V_{eff}=-\frac{E^2}{2} + \frac{1}{2}f(r)\left(\frac{J^2}{r^2}+1\right)
\end{equation}
where  $~~f(r)=1-\frac{2M}{r}+\frac{2M}{r}\tanh(\frac{Q^2}{2Mr})$.\\

The roots of $f(r)$ are the horizons. For radial geodesic(J=0), $V_{eff}$ will vanish for some finite value of r in the region $ 0 \leq r < r_{-}$ as $ f(r) > 0$. Therefore, it is not possible for timelike geodesic to reach the singularity. The massive particle will emerge in other region avoiding the singularity and the spacetime is geodesically complete. We can studied the motion for both $E=0$ and $E\neq 0$. In first case, if we take $E=0$ then $V_{eff}$ becomes
\[
V_{eff}=\frac{1}{2}\left(1-\frac{2M}{r}+\frac{2M}{r}\tanh\frac{Q^2}{2Mr}\right).
\]
The roots of $V_{eff}$ coincides with the horizons (Fig. 6) The shape of the potential indicates that the particle can move only inside the black hole. If we investigate the behaviour of $V_{eff}$ for $E \neq 0$, then when
 $ r\rightarrow 0$ the effective potential becomes
\begin{equation}
V_{eff} \rightarrow -\frac{2M}{r}+\frac{2M}{r}\tanh \frac{Q^2}{2Mr}\nonumber.
\end{equation}
Now, for large r,$V_{eff} \rightarrow \frac{1-E^2}{2}$. For a two horizons black hole, with in the two ranges $0 \leq r < r_{-} $ and $r_{+} < r$ , the function $f(r) > 0 $. Therefore, in these two region, it is possible for $V_{eff}$ to have roots. \\
Now consider for non-zero angular momentum($ J \neq 0$) particle. Roots of $V_{eff}$ coincides with the two horizons and the shape of $V_{eff}$ is shown in Fig.6. Therefore, we can say that the massive particle with zero energy never escape from black hole and would be in a bound orbit. As the potential has a minima, the particle may have circular stable orbit.\\
Again,for large r,$V_{eff} \rightarrow \frac{1-E^2}{2}$ when $E \neq 0$.
 When $ r\rightarrow 0$ then the effective potential becomes
\begin{equation}
V_{eff} \rightarrow \frac{J^2}{2r^2}\left(1-\frac{2M}{r}+\frac{2M}{r}\tanh \frac{Q^2}{2Mr}\right)\nonumber.
\end{equation}
The same conclusion can be taken for this case also that $V_{eff}$ have finite roots in between  $0 \leq r < r_{-} $ and $r_{+} < r$. Both the cases the massive particle would be in a bounded orbit.

\section{Solution of the Massless Scalar Wave Equation in the Charged E. Ay$\acute{o}$n-Beato and Garcia metric}
In this section we shall analyze the scalar wave equation for
charged E. Ay$\acute{o}$n-Beato and Garcia black hole geometry
following Brill et al \cite{Brill1972}. The wave equation for a
massless particle is given by
\begin{equation}
g^{-1/2} \frac{\partial}{\partial x^\mu} \left( g^{1/2} g^{\mu\nu} \frac{\partial}{\partial x^\nu} \right) \chi  =0
\end{equation}
Here, $g_{\mu\nu}$ is known from equation(1). Putting all the values one can obtain the following equation as
\begin{eqnarray}
-\frac{r^4 \sin\theta}{\Delta}\frac{\partial^2 \chi}{\partial t^2} + \sin \theta \frac{\partial}{\partial r}\left(\Delta \frac{\partial \chi}{\partial r}\right) \nonumber\\
+ \frac{\partial}{\partial \theta}\left( \sin\theta \frac{\partial}{\partial \theta} \right)\chi + \frac{1}{\sin\theta}\frac{\partial^2 \chi}{\partial \phi^2}=0 \nonumber
\end{eqnarray}
Where $\Delta = r^2 -2Mr+2Mr ~~tanh(\frac{Q^2}{2Mr})$.

Now, by using the separation of variable this equatin can be solved as
\[ \chi = e^{-i\omega t} e^{im\phi} R(r) \Theta(\theta) \]
Substuting this  we get
\begin{eqnarray}
\frac{r^4 \sin \theta}{\Delta}\omega^2 \chi + \frac{\sin \theta}{R} \frac{\partial}{\partial r}
\left(\Delta \frac{\partial R}{\partial r}\right) \chi \nonumber \\
+ \frac{1}{\Theta} \frac{\partial}{\partial\theta}\left(\sin\theta\frac{\partial\Theta}{\partial\theta} \right) \chi- \frac{m^2}{\sin\theta} \chi =0 \nonumber
\end{eqnarray}

Now,the radial equation reduces to
\[
\Delta \frac{\partial}{\partial r}\left( \Delta \frac{\partial R}{\partial r} \right) + (r^4 \omega^2 - \Delta \lambda)R=0
\]

and the angular part reduces to
\[
\frac{1}{\sin\theta}\frac{\partial}{\partial \theta} \left(\sin\theta\frac{\partial \Theta}{\partial\theta}\right) - \frac{m^2}{\sin^2\theta}\Theta + \lambda \Theta =0
\]
Now, if we substitute  $x=\cos \theta$, the equation becomes
\begin{equation}
\left(1-x^2\right)\frac{d^2 \Theta}{d x^2} - 2x \frac{d \Theta}{d x} - \left( \frac{m^2}{1-x^2}-\lambda \right) \Theta=0
\end{equation}
If we now write $\lambda = l(l+1)$ where  $l$ is an integer, then the equation
\[
\left(1-x^2\right)\frac{d^2 \Theta}{d x^2} - 2x \frac{d \Theta}{d x} + \left( l(l+1)-\frac{m^2}{1-x^2}  \right) \Theta=0
\]

is the familiar associated Legendre equation and the solution is given by the associated Legendre polynomial $ P^m_l(x)$ and is expressed as
\[
\Theta^m_l(\cos\theta) = P^m_l(x) = \frac{(-1)^m}{2^l l!}\left( 1-x^2 \right)^{m/2} \frac{d^{l+m}}{d x^{l+m}}(x^2-1)^l
\]

\subsection{The Radial Equation:Absence of Superradiance}

Kerr-Newman black hole allows energy extraction  whereas Schwarzschild black hole prohibits. By an explicit process, this can be achieved (Penrose Process).
Superradiance is nothing but the wave analogy of the Penrose process on black hole.When a bosonic or fermionic wave is incident on a black hole,
naturally the reflected wave carries less energy than the incident one. But under specific condition, the transmitted wave,absorbed by the black hole
carries negative energy into the black hole which makes the reflection coefficient greater than unity. This phenomena is
called 'superradiance'\cite{Misner1972}.According to this process the energy can be extracted from the black hole at the cost of lossing its
angular momentum.The requirred condition is \\
\begin{equation}
 0 < \omega < m \omega_H
\end{equation}
where $\omega_H $ is the angular velocity of the horizon \cite{DeWitt1975}. Basak et al discussed this phenomenon for the acoustic analogue of
the Kerr black hole \cite{Basak2003}. Shiraishi \cite{Shiraishi1992} and Ali \cite{Ali2007} argued that superradiance phenomenon is possible if the
black hole is rotating or charged.Now, we check whether the superradiance phenomenon really happens for E. Ay$\acute{o}$n-Beato and Garcia(ABG) black hole.\\
The radial equation can be written as
\[
\Delta \frac{d}{d r}\left( \Delta \frac{d R}{d r} \right) + \left( \omega^2 r^4 - l(l+1)\Delta\right)R=0
\]

Now, we introduce the familiar tortoise coordinate,$r^*$ defined as
\[
\frac{dr^*}{d r} = \frac{r^2}{\Delta}
\]
that implies
\[
\Delta \frac{d }{d r} = r^2 \frac{d }{d r^*}
\]

It is to be mentioned here that the variable $r^*$ is constructed in the same way as in Schwarzschild or in Kerr metric. Therefore, in this case the variable is non-integrable. Though the basic purpose is still satisfied, the coordinate spans over the real line and pushes the horizon to -$\infty$.\\
Again, we introduce  another function $u(r) = rR$ which reduces the radial equation in a more familiar form
\begin{eqnarray}
\frac{d^2 u}{d {r^*}^2}
 +[\frac{\Delta Q^2}{r^6}sech^2(\frac{Q^2}{2Mr})+\frac{2\Delta M}{r^5}(1-tanh(\frac{Q^2}{2Mr})) \nonumber\\
-\frac{2\Delta}{r^4}-\frac{l(l+1)\Delta}{r^4}+\frac{2\Delta^2}{r^6}+\omega^2 ] u = 0 \nonumber
\end{eqnarray}

Therfore, a potential barrier remains where

\begin{eqnarray}
V(r) = -[\frac{\Delta Q^2}{r^6}sech^2(\frac{Q^2}{2Mr})+\frac{2\Delta M}{r^5}(1-tanh(\frac{Q^2}{2Mr})) \nonumber\\
-\frac{2\Delta}{r^4}-\frac{l(l+1)\Delta}{r^4}+\frac{2\Delta^2}{r^6}+\omega^2 ]\nonumber
\end{eqnarray}

At horizon ($\Delta \rightarrow 0, r^* \rightarrow - \infty$), the radial equation comes out to be
\[
\frac{d^2 u_H}{dr{*2}} + \omega^2 u_H =0
\]
with $V(r) = -\omega^2$.

Now, asymptotically, $r \rightarrow \infty$, imlies that $r^* \rightarrow \infty$. The equation has the same form as in the previous case
\[
\frac{d^2 u_\infty}{dr{*2}} + \omega^2 u_\infty =0
\]

Thus $u_H = u_\infty$, where $u_H$ represents the radial solution
at horizon and $u_\infty$ is the solution at $\infty$. This
equality shows that for a  E. Ay$\acute{o}$n-Beato and Garcia
(ABG) black hole there is \textit{no phenomenon of superradiance
for an incident massless scalar field}.

\section{Conclusion}
In this investigation, we have analyzed the behavior of the
timelike and null geodesics of charged E. Ay$\acute{o}$n-Beato and
Garcia(ABG) Black hole. Here, we shown the behaviour of
time-distance and proper time-distance graph.In case of radial
geodesic, the effective potential is independent of the charge
and mass of the black hole, for photonlike particle. However,
from the shape of the potential, it is clear that the timelike
particle can move only inside the black hole. On the other hand,
for circular geodesics, the roots of the effective potential
coincide with the horizon. Also, from Fig. 5, we can say that as
the potential has a minima between the horizons, the photonlike
as well as timelike particles would be bounded in a stable
circular
orbit.\\
Though it is familiar that the superradiance phenomenon could be
seen in charged or rotating black
holes\cite{Shiraishi1992,Ali2007}, we have found that it is
absent in the charged E. Ay$\acute{o}$n-Beato and Garcia (ABG)
Black hole.

\section*{Acknowledgments} MK gratefully acknowledge support
 from IUCAA, Pune, India under Visiting Associateship under which a part
  of this work was carried out. SMH is  thankful to IUCAA also for giving him an opportunity to visit IUCAA where a part of this work was carried out.

\end{document}